\documentclass[]{article}
\usepackage{amsfonts,amssymb,amsmath,array}
\usepackage{graphicx}
\usepackage{amsbsy}
\usepackage{color}
\usepackage{hyperref} 
\usepackage{enumerate}
\usepackage{mathtools}
\usepackage{lipsum} 
\usepackage{cleveref} 
\usepackage[a4paper, total={7in, 10in}]{geometry}

\begin{document}

\title{Thermodynamic uncertainty relations in the presence of non-linear friction and memory}

\author{A. Plati$^1$ \and A. Puglisi$^{2,3,4}$ \and A. Sarracino$^5$}
\date{%
    $^1$Universit\'e Paris-Saclay, CNRS, Laboratoire de Physique des Solides, 91405 Orsay, France\\%
    $^2$Department of Physics, University of Rome Sapienza, P.le Aldo Moro 2, 00185, Rome, Italy\\
    $^3$Institute for Complex Systems - CNR, P.le Aldo Moro 2, 00185, Rome, Italy\\
    $^4$INFN, University of Rome Tor Vergata, Via della Ricerca Scientifica 1, 00133, Rome, Italy\\
    $^5$Department of Engineering, University of Campania ``Luigi Vanvitelli'', 81031 Aversa (CE), Italy \\
}


\maketitle

\begin{abstract}
A new Thermodynamic Uncertainty Relation (TUR) is derived for systems described by linearly coupled Langevin equations 
in the presence of non-linear frictional forces. In our scheme, the main  variable represents the velocity of a particle,
while the other coupled variables describe memory effects which may arise from strongly correlated degrees of freedom with several time-scales and, in general, are associated with thermal baths at different temperatures. The new TUR gives a lower bound for the mean-squared displacement of the position of the particle, including its asymptotic diffusion coefficient. This bound, in several examples worked out here, appears to be a good analytical estimate of the real diffusion coefficient. The new TUR can be also applied in the absence of any external force (with or without thermal equilibrium between the baths), a case which usually goes beyond the scope of original TURs. We show applications to non-linear frictional models with memory, such as the Coulomb and the Prantdtl-Tomlinson models, usually representative of friction at the nano-scale and within atomic-force microscopy experiments.
\end{abstract}

\section{Introduction}

In the last decade a family of rigorous inequalities has been derived, generally known as Thermodynamic Uncertainty Relations (TURs)~\cite{horowitz2020thermodynamic}. They provide bounds to fluctuations of several quantities, and those bounds are usually related to entropy production. The most simple and used result can be summarised as the entropy production rate limiting from above the precision rate of any non-equilibrium current of a driven system (a quantity similar but not exactly identical to a signal-to-noise ratio)~\cite{Barato2015,Gingrich2016,seifert2018stochastic,Hasegawa2019,Hasegawa2019II}. The TURs constitute a rare general result for out-of-equilibrium statistical mechanics, akin to Fluctuation-Dissipation Relations or Fluctuation Relations~\cite{marconi2008fluctuation}. They also provide an estimate of quantities which are not easily measurable, and this occurs in a twofold way: in certain cases entropy production is accessible/predictable, while the currents are not; in other cases the opposite happens~\cite{Hartich2021}. Of course, none of those problems is immediately solved by an inequality, but several examples have been given where the inequality is not far from being saturated~\cite{hwang2018energetic}. Moreover, a bound can be exploited through optimisation strategies. For instance, entropy production has been successfully approximated by machine-learning-based protocols, looking for the most precise non-equilibrium current which approaches the bound~\cite{kim2020learning}. 

In this paper we build upon a recently proposed inequality that relates entropy production and the mean-squared displacement (MSD) of a diffusing particle, valid also when in the presence of memory effects (e.g. degrees of freedom which are coupled to the particle velocity with different time-scales)~\cite{plati2023thermodynamic}. The main novelty in this work is represented by the possibility of considering also non-linear friction forces, i.e. dissipative forces that depend also non-linearly on the particle velocity~\cite{gennes2005brownian,hayakawa2005langevin,plyukhin2007nonlinear,baule2010stick,sarracino2013ratchet,manacorda2014coulomb,cerino2015entropy}. This category includes typical models such as the classical Coulomb friction, as well as power laws observed in foams~\cite{Cantat2013} and other non-linear models (such as the rate-and-state equation) that occur when considering microscopic and nanoscopic friction experiments~\cite{vanossi2013colloquium}. We will show how, in this framework, it is possible to derive analytically a tight lower bound for the diffusion coefficient, a quantity that requires the analytical form of the propagator to be computed exactly \cite{Lequy2023}.  

The paper is organized as follows.
In Section 2 we discuss a general strategy to derive TURs which is different from previously proposed approaches, and which is particularly useful in our case. In Section 3, we apply such a derivation to the problem of generalised Langevin equations with exponential memory kernel and non-linear frictional velocity-dependent forces. This class of systems can be mapped - by Markovian embedding - into linearly coupled Markovian Langevin equations, still maintaining the non-linear forces in the equation for the particle motion. In Section 4, we draw conclusions and propose future perspectives.

\section{General strategies for the derivation of TURs}
\label{sec:derivation}

The TURs have been initially conjectured and then derived, by means of several techniques, for a larger and larger category of systems, encompassing Markovian discrete and continuous dynamics as well as transient and steady states~\cite{Barato2015,Gingrich2016,dieball2023direct}. 
In recent years, a general strategy to derive TURs-like inequalities, including bounds for linear response, has been traced in the framework of information theory~\cite{dechant2018multidimensional}. It goes through the introduction of a virtual perturbation in the original model equations, and then the application of the Cram\'er-Rao inequality, which identifies the reciprocal of the Fisher Information as a lower bound for the variance of an unbiased estimator~\cite{steven1993fundamentals}. The usual application of this idea is introducing the virtual perturbation in the equations with the aim of modifying the irreversible probability current of the system, that is the part of the probability current which is responsible for the lack of detailed balance (and is therefore connected to entropy production). However, such a strategy cannot be directly applied to a system with underdamped dynamics, raising the necessity of alternative treatments~\cite{Hasegawa2019II}. Recently, we have proposed a different application of the Cram\'er-Rao-based derivation, in the context of Stochastic Differential Equations (SDE): the virtual perturbation is applied directly to the variables appearing in the SDE, and the symmetries of the system are exploited to have additional contributions that can be optimised to obtain tighter bounds~\cite{plati2023thermodynamic}. Here we build upon such an approach to include a non-linear friction-force term in the main (velocity) variable.

\subsection{General bound in the presence of non-linear friction and memory}

We consider a Generalized Langevin Equation with exponential memory kernel~\cite{loos2021stochastic} for a particle with position $\theta$ and velocity $\omega$, subjected to a nonlinear force $\mathcal{F}(\omega)$  and dragged by a constant force $F_{\text{ext}}$~\cite{puglisi2009irreversible}:

\begin{eqnarray}
\dot{\theta}&=&\omega,\\
 \dot{\omega}(t)&=&-\int_{-\infty}^t\gamma(t-t')\omega(t') dt'+\eta_s(t) +F_{\text{ext}}
+\mathcal{F}(\omega),\label{GLE} \\
 \gamma(t)&=&\frac{2}{\tau}\delta(t)+\sum_{k=1}^n a_k e^{-\frac{t-t'}{\tau_k}},\\
 \langle \eta_s(t)\eta_s(t') \rangle&=&\frac{2q}{\tau}\delta(|t-t'|)+ \sum_{k=1}^n q_k a_k e^{-\frac{|t-t'|}{\tau_k}}.
\end{eqnarray}
Upon defining the $n$ auxiliary variables:

\begin{equation}
\Omega_k=-b_k\int_{-\infty}^t dt' e^{-\frac{t-t'}{\tau_k}}\left[a_k\omega(t')-\sqrt{\frac{2q_k a_k}{\tau_k}}\xi_k(t')\right]  \quad k \in [1,n],
\end{equation}
the system can be written as a Markovian SDE for the $n+1$ dimensional vector $\boldsymbol{X}=\{\omega,\Omega_1,\ldots,\Omega_n\}$:

\begin{subequations}\label{eq::SDEnl}
\begin{equation}
\dot{\omega}=-\frac{\omega}{\tau}+\sum_k\frac{\Omega_k}{b_k}+\sqrt{\frac{2q}{\tau}}\xi(t)+ F_{\text{ext}}+\mathcal{F}(\omega),
\end{equation}
\begin{equation}
\dot{\Omega}_k=-\frac{\Omega_k}{\tau_k}-a_k b_k\omega+\sqrt{\frac{2q_k a_k b_k^2}{\tau_k}}\xi_k(t).
\end{equation}
\end{subequations}

We define accordingly the noise amplitudes $B_i$, such as $B_0=\sqrt{2q/\tau}$
and $B_i=\sqrt{2q_i a_i b_i^2/\tau_i}$ for $i>0$. From Eqs. \eqref{eq::SDEnl}  we obtain the following relation between the average values of $\omega$ and $\Omega_k$s in the stationary state:

\begin{equation}\label{eq::Meannl}
  \quad
  \langle \Omega_k\rangle = -\tau_k a_k b_k \langle \omega\rangle.
\end{equation}

The strategy to derive the new inequality is to introduce a perturbation in Eqs. \eqref{eq::SDEnl} by adding to the r.h.s. of each equation the corresponding component of the vector $h\boldsymbol{V}(\omega)$, which is defined as:
\begin{equation}
V_0(\omega)=\langle \omega\rangle/\tau - \langle \omega\rangle\mathcal{F}'(\omega) -\sum_k u_k/b_k \quad \text{and} \quad V_k=a_k b_k\langle \omega\rangle + u_k/\tau_k,
\end{equation}
where $u_k$ are arbitrary parameters, to be fixed later.

Considering that the $h$-perturbation can be arbitrarily small, we are allowed to take $\mathcal{F}(\omega)-h\langle\omega\rangle\mathcal{F}'(\omega)\simeq \mathcal{F}(\omega-h\langle\omega\rangle)$.
Then, the perturbed dynamics is equivalent to the dynamics described by Eqs. \eqref{eq::SDEnl} for the vector $\boldsymbol{X}_h=\{\omega -h\langle \omega \rangle,\Omega_1 - hu_1,\ldots,\Omega_n-hu_n\}$. In other words, the perturbed $h-$dynamics is the same as the original one, but with shifted averaged values:
\begin{equation}
\langle \omega \rangle_h=(1+h)\langle \omega \rangle, \quad \langle \Omega_k \rangle_h=\langle \Omega_k \rangle+hu_k,
\end{equation}
and for the perturbed steady state we have $P_h(\boldsymbol{X})=P(\boldsymbol{X}_h)$, where $P_h$ and $P$ are the stationary distributions for the perturbed and unperturbed dynamics, respectively. 
Now we apply the Cram\'er-Rao inequality that, in our case, reads:
\begin{equation}\label{eq::CramRao}
 \frac{\text{Var}_{h=0}(\theta(t))}{[\partial_{h=0} \langle \theta(t) \rangle_{h}]^2} = \frac{\langle \Delta\theta(t)^2 \rangle}{(\langle \omega \rangle t)^2} \ge \frac{1}{\mathcal{I}_{\text{F}}(h=0)}, 
 \end{equation} 
where $\Delta \theta(t)=\theta(t) - \langle \theta\rangle$,
\begin{equation}
\mathcal{I_{\text{F}}}(h=0)=-\langle \partial^2_h \ln P_h(\boldsymbol{X}) \rangle_{h=0} + \left\langle \int_0^t dt'\sum_i \left( \frac{V_i(\omega)}{B_{i}}   \right)^2 \right\rangle_{h=0}=\mathcal{I}/2+\mathcal{B}t,
\end{equation} 
and 
\begin{equation}\label{eq::defIandB}
\mathcal{I}=2\int d\boldsymbol{X} \frac{[\partial_h P_h(\boldsymbol{X})]^2|_{h=0}}{P( \boldsymbol{X})}, \quad \mathcal{B}=\left(\frac{\langle \omega \rangle}{\tau} -   \langle \omega\rangle \langle\mathcal{F}'\rangle -\sum_j\frac{u_j}{b_j} \right)^2 B_{0}^{-2}+\sum_k\left(a_kb_k\langle \omega \rangle +\frac{u_k}{\tau_k}\right)^2B_{k}^{-2}.
\end{equation}
Note that $\langle\mathcal{F}'\rangle$ requires the knowledge of the stationary distribution  $P(\boldsymbol{X}_h)$ to be computed.
The values of $u_k$ as a function of the model parameters can be fixed by minimising $\mathcal{B}(\boldsymbol{u})$. In this way, one obtains a bound which is optimal in this set of possible perturbations. This is done by solving the system of equations defined by $\partial \mathcal{B}(\boldsymbol{u})/\partial u_k=0
$ $\forall k$. The system is formally solved by $\boldsymbol{u}=\hat{T}^{-1}\boldsymbol{z}$ where:
\begin{equation}\label{eq::SystForu}
T_{ii}=\frac{1}{\tau_i^2B_{i}^2}+\frac{1}{b_i^2B_{0}^2},\quad T_{i\neq j}=\frac{1}{b_ib_j B_{0}^2}, \quad z_i=\langle\omega\rangle\left[\frac{\tau^{-1}+\langle\mathcal{F}'\rangle}{b_i B_{0}^2}-\frac{a_i b_i}{\tau_iB_{i}^2} \right].
\end{equation}
First, we note that with equal thermostats $q_i=q$ $\forall i$   and $\langle\mathcal{F}'\rangle=0$, one has  $z_i=0$, which means that the optimal solution is  $u_k=0$ $\forall$ $k$. This is consistent with the analytical result for the linear system where the bound for the diffusion coefficient is saturated at equilibrium. 

In order to find the analytical expression of the set of values $\{u_k\}$, an exact inversion of the matrix $\hat{T}$ is required, leading to very cumbersome formulae. However, from Eq. \eqref{eq::SystForu}, one has that the solutions
can be written as $u_k=\langle \omega\rangle\tilde{u}_k$. It is important to note that this is not an assumption of linearity in $\langle \omega\rangle$, but a simple rescaling that will allow us to obtain some simplifications in the final inequality.

Now we can define $\mathcal{I}=\langle \omega\rangle^2\tilde{\mathcal{I}}$ and $\mathcal{B}=\langle \omega\rangle^2\tilde{\mathcal{B}}$ and simplify $\langle \omega\rangle^2$ in the bound for the msd Eq. \eqref{eq::CramRao}  obtaining:
\begin{equation} \label{eq::nlBound}
\langle \Delta\theta(t)^2 \rangle \ge \frac{t^2}{\tilde{\mathcal{B}}t+\tilde{\mathcal{I}}/2}.
\end{equation}
We note that, from Eq. \eqref{eq::defIandB},
it is easy to verify that $\mathcal I$ is proportional to $\langle \omega\rangle^2$.
Note that from the equation above and Eqs. \eqref{eq::defIandB}, one has that the case without memory ($a_k=u_k=0$ $\forall$ $k$) could lead to the pathological condition $1/\tau-\langle \mathcal{F}' \rangle=0$ for which Eq. \eqref{eq::nlBound} predicts a non-physical asymptotic ballistic regime. However, for frictional forces, which constitute the main focus of this paper, $\mathcal{F}'$ is non-positive, thus preventing any irregularity of the bound. Anyway, even when $\mathcal{F}'\ge 0$, the system without memory admits a potential solution so that the quantity $1/\tau-\langle \mathcal{F}' \rangle$ can be evaluated analytically to check the validity of Eq. \eqref{eq::nlBound}.
Moreover, from Eqs. \eqref{eq::defIandB} and \eqref{eq::nlBound}, one can also check that, in the non-optimal case ($u_k=0$ $\forall$ $k$) and without nonlinear friction ($\mathcal{F}(\omega)=0$), the TUR derived in Ref.~\cite{plati2023thermodynamic} for linear forces is recovered. In this condition, the specific contribution to the entropy production rate due to the presence of external forcing appears in the denominator of the bound.

Both $\tilde{\mathcal{B}}$ and $\tilde{\mathcal{I}}$ are not known in general, since they require the knowledge of the stationary distribution $P(\boldsymbol{X})$. In the following, we will concentrate on cases where the r.h.s. of Eq. \eqref{eq::nlBound} can be found analytically.         

\subsection{Two classes of models}

Equation \eqref{eq::nlBound} can be used as an analytical bound for the diffusion coefficient 
in the following classes of models: i) models with no memory, featuring generic nonlinear friction;  
ii) models with memory, including nonlinear friction with a bounded derivative, in the large time limit.
In the first case,  we have a gradient system for a single degree of freedom. We can then find the stationary probability distribution and in turn compute $\tilde{\mathcal{B}}$ and $\tilde{\mathcal{I}}$.
In the second case, exploiting the maximum of the modulus of the derivative of the non-linear force $\max(|\mathcal{F}'|)$, we can use the following chain of inequalities
\begin{equation}\label{eq::chainIneq}
\left(\frac{\langle \omega \rangle}{\tau} -   \langle \omega\rangle \langle\mathcal{F}'\rangle -\sum_j\frac{u_j}{b_j} \right)^2 \le \left(\frac{\langle \omega \rangle}{\tau} +   \langle \omega\rangle |\langle\mathcal{F}'\rangle| -\sum_j\frac{u_j}{b_j} \right)^2 \le \left(\frac{\langle \omega \rangle}{\tau} +   \langle \omega\rangle  \max(|\mathcal{F}'|)  -\sum_j\frac{u_j}{b_j} \right)^2,
\end{equation}
and define 
\begin{equation}
\mathcal{B}_{\text{max}}=\left(\frac{\langle \omega \rangle}{\tau} +   \langle \omega\rangle  \max(|\mathcal{F}'|)  -\sum_j\frac{u_j}{b_j} \right)^2B_{0}^{-2}+\sum_k\left(a_kb_k\langle \omega \rangle +\frac{u_k}{\tau_k}\right)^2B_{k}^{-2}\ge\mathcal{B},
\end{equation}
to have an approximate but analytical bound for the MSD at long times (i.e. for the diffusion coefficient):
\begin{equation}
\lim_{t\to \infty} \frac{1}{t}\langle \Delta\theta(t)^2 \rangle \ge \frac{1}{\tilde{\mathcal{B}}_{\text{max}}},
\end{equation}
where $\tilde{\mathcal{B}}_{\text{max}}=\mathcal{B}_{\text{max}}/\langle \omega \rangle^2$. This is an explicit bound with a closed analytical form. It is however weaker than the non-analytic one in which $\langle\mathcal{F}'\rangle$ explicitly appears. Nevertheless, it can still be improved by choosing the optimal $u_k$ that minimize $\tilde{\mathcal{B}}_{\text{max}}(\boldsymbol{u})$  (note that the chain of inequalities \eqref{eq::chainIneq} is written for generic $u_k$).

\subsection{Comparison with the standard TUR}
For a large class of stochastic processes, which include the model Eqs. \eqref{eq::SDEnl} under study, a bound for asymptotic diffusion can be derived through the standard TUR involving the entropy production rate (EPR) $\langle\dot{S}\rangle$ in the steady state \cite{horowitz2020thermodynamic,plati2023thermodynamic}:
\begin{equation}\label{eq::StandardTURbound}
\lim_{t\to \infty} \frac{1}{t}\langle \Delta\theta(t)^2 \rangle\ge \frac{2\langle \omega \rangle^2}{ \langle\dot{S}\rangle}.
\end{equation}
Here, we compare this previous result with the bound Eq. \eqref{eq::nlBound} derived in this paper. 
The EPR in the steady state for our model can be easily computed by splitting the non-linear velocity-dependent force into a reversible and irreversible part: $\mathcal{F}(\omega)=\mathcal{F}_{\text{rev}}(\omega)+\mathcal{F}_{\text{irr}}(\omega)$, where $\mathcal{F}_{\text{rev}}(\omega)=1/2 [\mathcal{F}(\omega)+\mathcal{F}(-\omega)]$ and $\mathcal{F}_{\text{irr}}(\omega)=1/2 [\mathcal{F}(\omega)-\mathcal{F}(-\omega)]$ . Then, using the results in~\cite{puglisi2009irreversible}, we obtain:
\begin{equation}
\begin{split}
\langle\dot{S}\rangle=&\frac{1}{q}\langle\omega F_{\text{ext}}\rangle+\frac{1}{q}\langle\omega\mathcal{F}_{\text{rev}}(\omega)\rangle-\langle \omega \rangle^2\sum_ia_i\tau_i\left(\frac{1}{q} -\frac{1}{q_i}\right)+\sum_i\frac{1}{b_i}\left(\frac{1}{q} -\frac{1}{q_i}\right)\sigma_{0i} \\& -\frac{\tau}{q}\sum_i\frac{1}{b_i}\langle \mathcal{F}_{\text{irr}}(\omega)\Omega_i \rangle-\frac{\tau}{q}\langle \mathcal{F}_{\text{irr}}(\omega)F_{\text{ext}} \rangle-\frac{\tau}{q}\langle \mathcal{F}_{\text{irr}}(\omega)\mathcal{F}_{\text{rev}}(\omega) \rangle,
\end{split}
\end{equation}
where $\sigma_{0i}=\langle (\omega-\langle\omega\rangle)(\Omega_i-\langle\Omega_i\rangle) \rangle$.

In the cases of interest for this paper, the velocity-dependent force has only the irreversible part ($\mathcal{F}_{\text{rev}}(\omega)=0$) and the expression substantially simplifies: 
\begin{equation}\label{eq::StandardTURmem}
 \langle\dot{S}\rangle=\frac{1}{q}F_{\text{ext}}\langle\omega\rangle-\langle \omega \rangle^2\sum_ia_i\tau_i\left(\frac{1}{q} -\frac{1}{q_i}\right) +\sum_i\frac{1}{b_i}\left(\frac{1}{q} -\frac{1}{q_i}\right)\sigma_{0i}-\frac{\tau}{q}\sum_i\frac{1}{b_i}\langle \mathcal{F}_{\text{irr}}(\omega)\Omega_i \rangle-\frac{\tau}{q}F_{\text{ext}}\langle \mathcal{F}_{\text{irr}}(\omega) \rangle,
 \end{equation} 
where $\mathcal{F}_{\text{irr}}(\omega)$ refers to any frictional force.
The case without memory is simpler:
\begin{equation}\label{eq::StandardTURnomem}
 \langle\dot{S}\rangle=\frac{1}{q} F_{\text{ext}}\langle\omega\rangle -\frac{\tau}{q}F_{\text{ext}} \langle \mathcal{F}_{\text{irr}}(\omega)\rangle.
 \end{equation} 
As we expect from~\cite{sarracino2013time}, the dynamics is reversible when there is a single thermostat and $F_{\text{ext}}=0$. Note that in the presence of a spatial potential, the system would have been out-of-equilibrium also in the case $F_{\text{ext}}=0$~\cite{sarracino2013time}.

The bound predicted by the standard TUR coincides with Eq. \eqref{eq::StandardTURbound} with $\langle\dot{S}\rangle$ given by Eqs. \eqref{eq::StandardTURmem} and \eqref{eq::StandardTURnomem}.
However, there are many crucial differences between this bound and Eq. \eqref{eq::nlBound}: 

\begin{enumerate}
\item The standard bound is valid only in the long-time regime.
\item It requires the computation of non-trivial averages on the steady state, that are more difficult to bound with respect to $\langle\mathcal{F}' \rangle$. Thus, making the case with memory analytical is non-trivial. 
\item In general, its denominator does not scale with $\langle \omega^2 \rangle$ and therefore the case of free diffusion (i.e. without any applied force) becomes more delicate to treat; in our result, on the contrary, the scaling with $\langle \omega^2 \rangle$ allows the free diffusion limit to have a well defined bound.
\end{enumerate}
From the above remarks, it follows that a connection with our bound is not a direct task for the case with $F_ {ext}=0$. A comparison can be done (see Section \ref{sec::GradCase}) in the absence of memory: exploiting the potential form of the solution of the gradient system, one can explicitly compute the averages and check if a good scaling of $\langle \mathcal{F}_{\text{irr}}(\omega)\rangle$ with $\langle \omega\rangle$ occurs. In general, the bound derived from our procedure is more appropriate for the estimate of the diffusion coefficient.

\section{Applications}
\label{sec:applications}

In this Section, we apply our results to specific models. In particular, we consider two relevant kinds of non-linear velocity-dependent forces $\mathcal{F}(\omega)$: the Coulomb and the Prandtl-Tomlinson (PT) friction models. For each of them, we study a case without memory and a case with memory with a single characteristic time. The class of considered systems is defined by the general model:
\begin{subequations}\label{eq::SDEnl_numTest}
\begin{equation}\label{eq::numTestOmega}
\dot{\omega}=-\frac{\omega}{\tau}+\frac{\Omega}{b}+\sqrt{\frac{2q}{\tau}}\xi(t)+ F_{\text{ext}}+\mathcal{F}(\omega),
\end{equation}
\begin{equation}
\dot{\Omega}=-\frac{\Omega}{\tau_1}-a b\omega+\sqrt{\frac{2q_1 a b^2}{\tau_1}}\xi_1(t),
\end{equation}
\end{subequations}
where the limit of no memory, in the stationary state, corresponds to $a=0$.

In order to have an explicit expression for the bound, we need to solve the equation for the optimal $u$ which in this case is a scalar. Of course, we have to do this optimization for $\mathcal{B}_{\text{max}}$ and not for $\mathcal{B}$ so that the optimal $u$ will be a function of $\max(|\mathcal{F}'|)$ and not of $\langle \mathcal{F}' \rangle$:
\begin{equation}\label{eq::optu}
\tilde{u}=u/\langle\omega\rangle=T^{-1}z/\langle\omega\rangle=\left(\frac{\tau^{-1}+\max(|\mathcal{F}'|)}{b B_{0}^2}-\frac{ab}{\tau_1B_{1}^2} \right)\left(\frac{1}{\tau_1^2B_{1}^2}+\frac{1}{b^2B_{0}^2}\right)^{-1}.
\end{equation}
We now define the non-linear forces of interest:
\begin{equation}
\mathcal{F}_{\text{Coulomb}}(\omega)=\begin{cases}-\omega/\tau_\mu \quad \text{if} \quad |\omega|\le\mu \\ -(\mu/\tau_\mu)\textrm{sign}(\omega) \quad \text{if} \quad |\omega|>\mu,  \end{cases}\label{eq::Coulomb}
\end{equation}
which describes the Coulomb friction law with a viscous regime for small velocities $|\omega|\le\mu$. The latter is generally needed both for numerical reasons (i.e. to avoid discontinuous forces) but it has also a physical motivation in frictional contact models. 

The PT friction force reads:
\begin{equation}
\mathcal{F}_{\text{PT}}(\omega)=\begin{cases} -\omega/\tau_{\mu} \quad \text{if} \quad |\omega|\le\mu \\ -(\mu/\tau_\mu)\textrm{sign}(\omega)\left[1+\ln (|\omega|/\mu) \right]  \quad \text{if} \quad |\omega|>\mu.  \end{cases} \label{eq::Fpt} 
\end{equation}
In both models, $\tau_\mu$ and $\mu$ are positive constants. Note that Eq. \eqref{eq::Fpt} coincides with Eq. (1) of \cite{Muser2011}, with the constants defined in such a way as to have a continuous function with a continuous derivative. 

In order to have an explicit expression of the bound we need the maximum modulus of the non-linear forces, which is in both cases $1/\tau_\mu$
\begin{equation}\label{eq::maxMods}
\max(|\mathcal{F}_{\text{Coulomb}}'|)=\frac{1}{\tau_\mu}, 
\quad \max(|\mathcal{F}_{\text{PT}}'|)=\frac{1}{\tau_\mu}.
\end{equation}
We have the following bounds for the cases with memory:
\begin{align}\label{eq::numCheckValidity}
\lim_{t\to \infty} \frac{1}{t}\langle \Delta\theta(t)^2 \rangle &\ge \frac{1}{\tilde{\mathcal{B}}_{\text{max}}} \ge \frac{1}{\tilde{\mathcal{B}}_{\text{max,2}}},\\
\tilde{\mathcal{B}}_{\text{max}}&=\left(\frac{1}{\tau} +  \max(|\mathcal{F}'|)  -\frac{\tilde{u}}{b} \right)^{2}B_{0}^{-2}+\left(ab+\frac{\tilde{u}}{\tau_1}\right)^2B_{1}^{-2}, \\
\tilde{\mathcal{B}}_{\text{max,2}}&=\left(\frac{1}{\tau} +  \max(|\mathcal{F}'|) \right)^{2}B_{0}^{-2}+\left(ab\right)^2B_{1}^{-2},
\end{align}
where $\max(|\mathcal{F}'|) $ is defined in Eq. \eqref{eq::maxMods} for the two forces, and $\tilde{u}$ is given in Eq. \eqref{eq::optu}. The first inequality in Eq.~\eqref{eq::numCheckValidity} represents one of our main results, while the second one is considered to show how the optimized $\tilde{u}$ can substantially improve the bound.

\subsection{Coulomb friction with and without memory}

\begin{figure}
\centering
\includegraphics[width=0.5\columnwidth,clip=true]{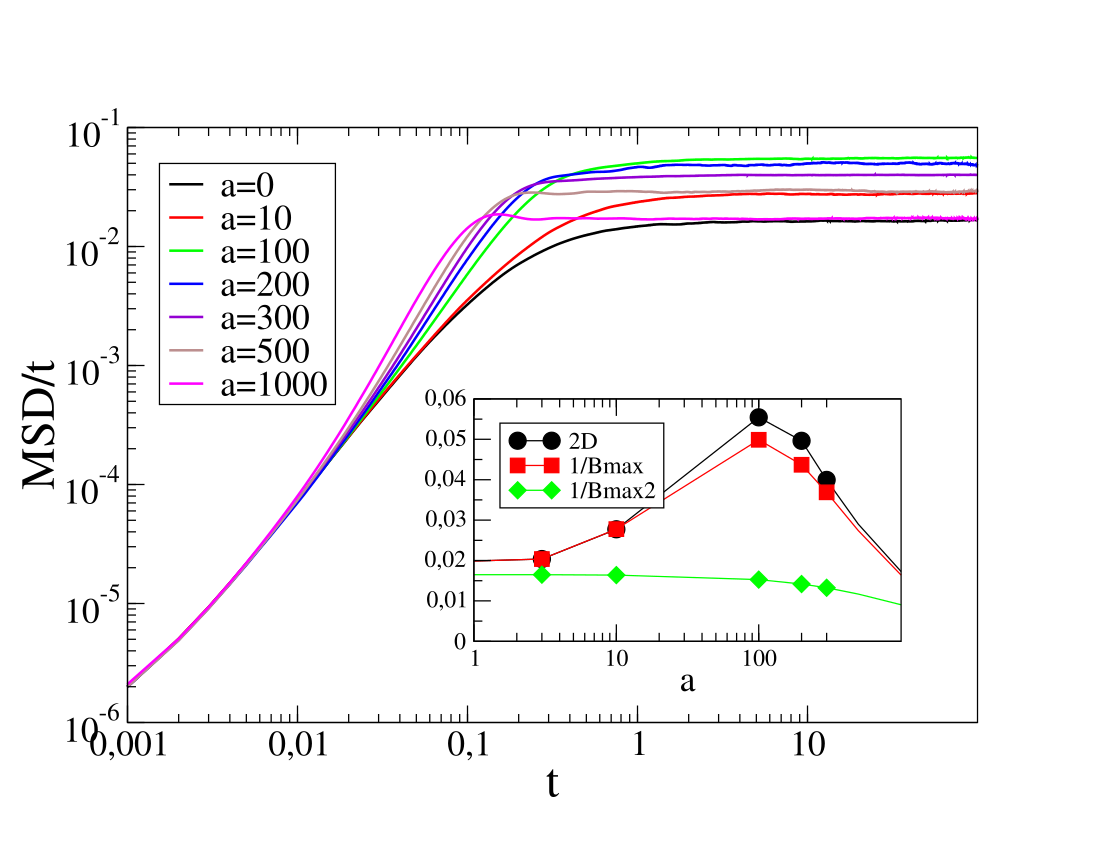}
\caption{MSD, diffusion coefficient and analytical bounds for the Coulomb friction model with $q=q_1=1$. Other parameters are: $\tau=1$, $b=0.01$, $F_{\text{ext}}=0.1$, $\tau_\mu=0.1$, $\mu=1$, $\tau_1=0.1$. Here and in the following, we measure time in $\tau$ units. The integration time-step used in the numerical simulations is $dt=10^{-3}$. \label{fig:Fig1}}\end{figure}

In Fig.~\ref{fig:Fig1} we show the case where both variables are at temperature $q=q_1=1$, i.e. the system is at equilibrium unless the external force is switched on. 
Several curves in the main graph illustrate the behavior of the mean squared displacement (rescaled by time) for different values of the memory coupling $a$, where we remind that $a=0$ corresponds to a lack of memory. In the inset (black dots), we consider the diffusion coefficient $2D=\lim_{t\to\infty}\langle [\theta(t)-\theta(0)]^2\rangle/t$, as a function of $a$. The first observation is the non-monotonic behaviour of the diffusion coefficient as a function of $a$. In the descending regime (the values of $a$ for which $D$ decreases with $a$) the curve of $MSD/t$ appears to be also non-monotonous. Apparently, the second time-scale introduced by the memory is responsible for a kind of dynamical slowing down which eventually influences the asymptotic diffusion, reducing $D$. In the inset, we also report the two bounds $1/B_{\text{max}}$ and $1/B_{\text{max,2}}$. The tightest bound fairly approximates the diffusion coefficient, including its non-monotonic behaviour. The non-optimised bound, on the contrary, gets close to $D$ only in the limit $a \to 0$ (no memory), while in the rest of the range it significantly underestimates $D$, showing a monotonic trend.

\begin{figure}
\centering
\includegraphics[width=0.5\columnwidth,clip=true]{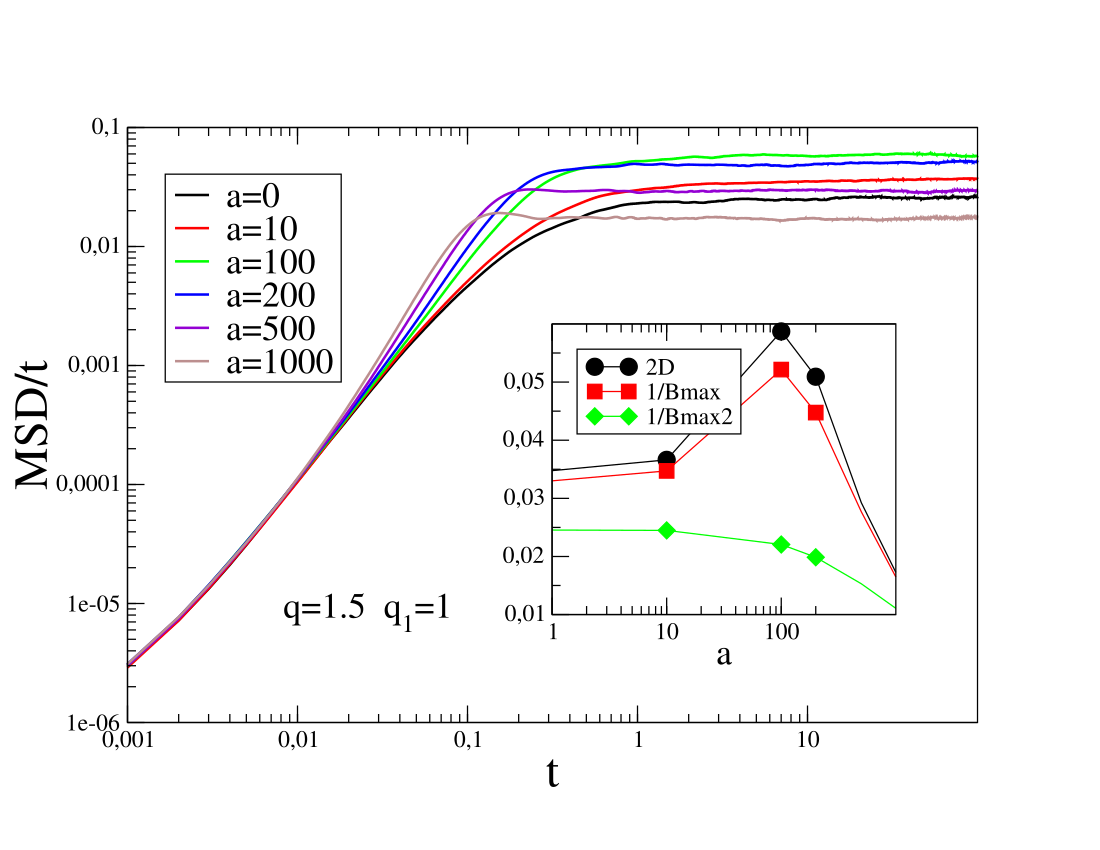}
\caption{MSD, diffusion coefficient and analytical bounds for the Coulomb friction model with $q=1.5$ and $q_1=1$. Other parameters as in Fig.~\ref{fig:Fig1}.  \label{fig:Fig2}}\end{figure}

In Fig.~\ref{fig:Fig2} we repeat the same analysis in a weakly out-of-equilibrium case $q=1.5$ and $q=1$ (the variable representing memory is slightly hotter than the main velocity variable). The scenario looks similar to the equilibrium case, with a non-monotonic behavior of both $MSD/t$ vs. $t$ (for large $a$) and of $D$ vs $a$. In the inset, we report the two bounds $1/B_{\text{max}}$ and $1/B_{\text{max,2}}$, finding the same behaviour observed in the previous case.

\begin{figure}
\centering
\includegraphics[width=0.5\columnwidth,clip=true]{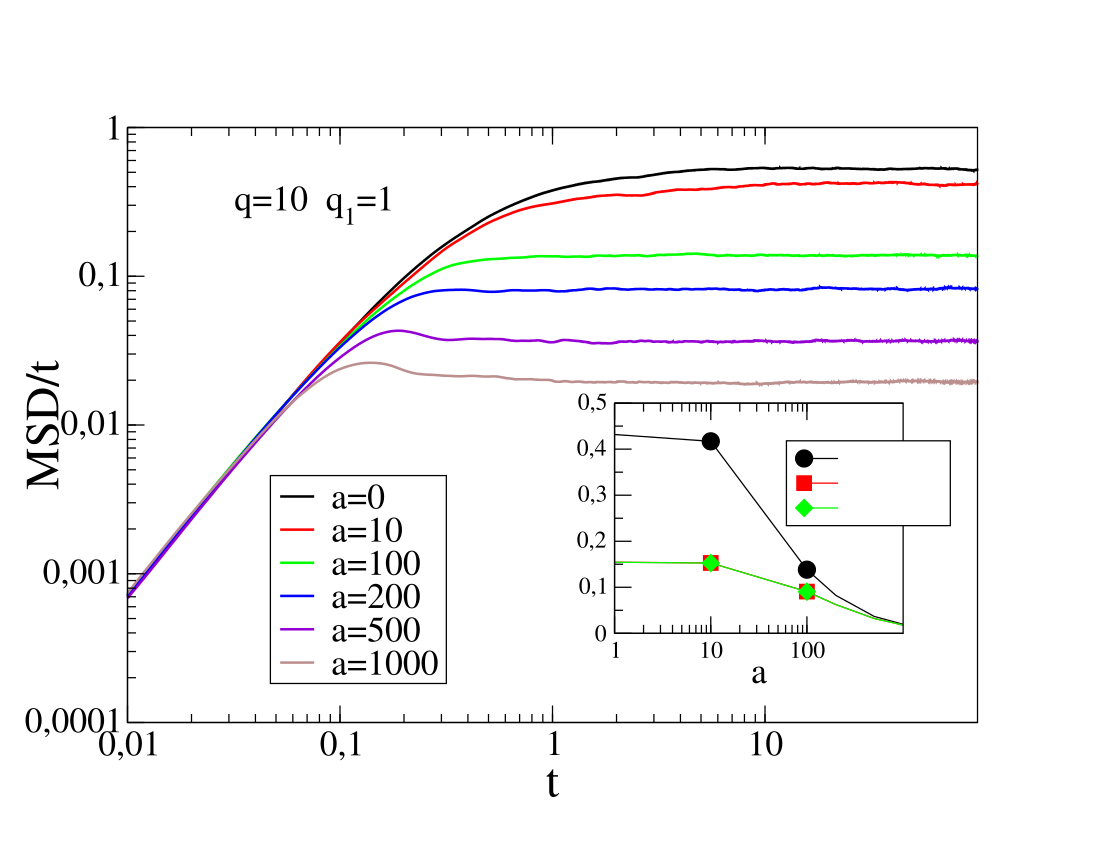}
\caption{MSD, diffusion coefficient and analytical bounds for the Coulomb friction model with $q=10$ and $q_1=1$.   Other parameters as in Fig.~\ref{fig:Fig1}.\label{fig:Fig3}}\end{figure}

Last, we consider a strongly out-of-equilibrium case  with $q=10$ and $q_1=1$, in Fig.~\ref{fig:Fig3}. Here, while the curves of $MSD/t$ vs. $t$ are still non-monotonous when memory is strong (large $a$), the scenario for $D$ vs. $a$ is different from the previous cases, as it is monotonously decreasing. The bound is confirmed and also monotonously decreasing, but it heavily underestimates $D$ in the limit of weak memory.

\subsection{Prandtl-Thomlinson friction with and without Memory.}

\begin{figure}
\centering
\includegraphics[width=0.5\columnwidth,clip=true]{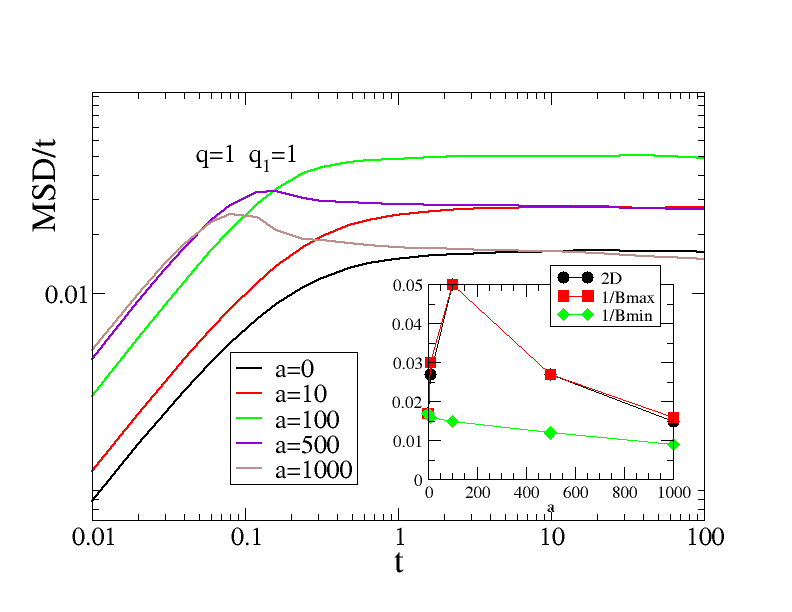}
\caption{MSD, diffusion coefficient and analytical bounds for the PT friction model with $q=q_1=1$. Other parameters are: $\tau=1$, $b=0.01$, $F_{\text{ext}}=0.1$, $\tau_\mu=0.1$, $\mu=1$, $\tau_1=0.1$. \label{fig:Fig4}}\end{figure}

\begin{figure}
\centering
\includegraphics[width=0.5\columnwidth,clip=true]{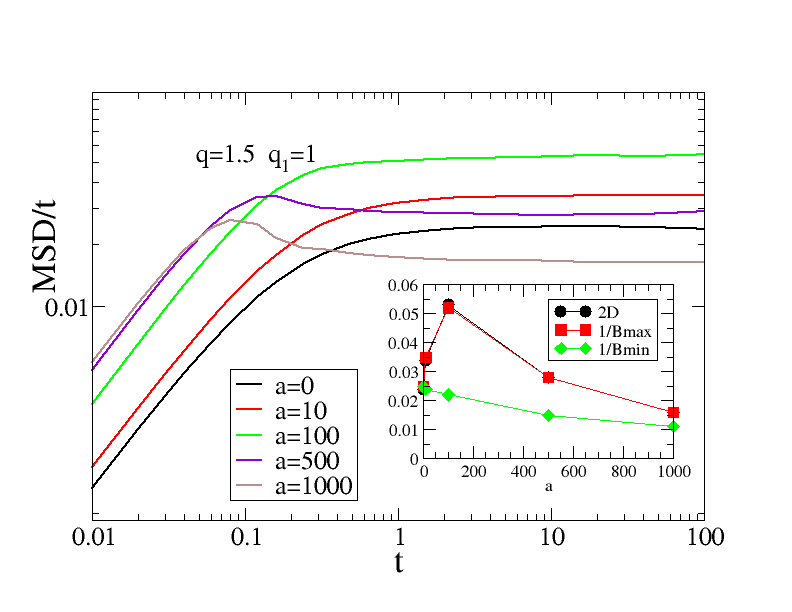}
\caption{MSD, diffusion coefficient and analytical bounds for the PT friction model with $q=1.5$ and $q_1=1$. Other parameters as in Fig.~\ref{fig:Fig4} \label{fig:Fig5}}\end{figure}

\begin{figure}
\centering
\includegraphics[width=0.5\columnwidth,clip=true]{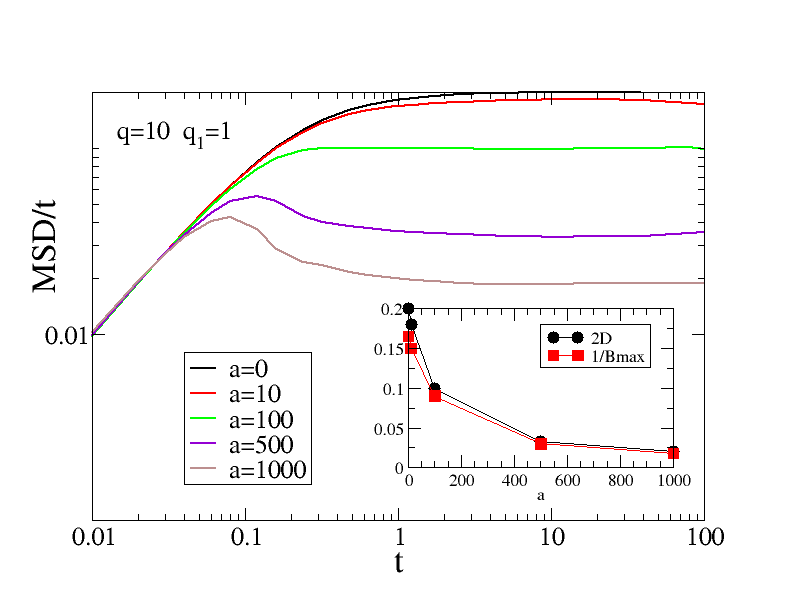}
\caption{MSD, diffusion coefficient and analytical bounds for the PT friction model with $q=10$ and $q_1=1$. Other parameters as in Fig.~\ref{fig:Fig4}\label{fig:Fig6}}\end{figure}

In this Section, we study the PT frictional model. Figures~\ref{fig:Fig4},~\ref{fig:Fig5} and~\ref{fig:Fig6} show the results of simulations for equilibrium, weakly out-of-equilibrium and strongly out-of-equilibrium cases, respectively. The scenario is quite similar to that observed in the Coulomb model. At large memory the $MSD/t$ vs $t$ curve is non-monotonous, and $D$ vs. $a$ is also non-monotonous in the equilibrium and weakly out-of-equilibrium regimes. In these two regimes the optimised bound works very well, and much better than the non-optimised case. In the strongly out-of-equilibrium case, the non-monotonicy of $D$ vs. $a$ disappears and the two bounds become indistinguishable. Remarkably, for this model, the bound provides a nice approximation of $D$ for a much larger range of memory coupling values $a$, with respect to the Coulomb case.

\subsection{Discussion of analytical results for the gradient cases} \label{sec::GradCase}

The non-linear friction models considered here are, usually, not analytically solvable. Nevertheless, in the absence of memory, the single variable stochastic differential equation becomes a gradient system and this gives easily an expression for the steady-state probability distribution:
\begin{equation}\label{eq::Pomega}
    P(\omega)=\frac{\mathcal{N}\tau}{2q}\exp \left[\frac{1}{q} \int_0^\omega d\omega' \left(-\omega'+\tau F_{\text{ext}}+\tau\mathcal{F}_{\text{Cou}}(\omega')\right) \right],
\end{equation}
where $\mathcal{N}$ is the normalization factor.
The above formula is useful to get analytical results for the quantities $\mathcal{B}$ and $\mathcal{I}$ which are included in the most general bound expression, Eq.~\eqref{eq::nlBound}. 

\begin{figure}
\centering
\includegraphics[width=0.4\columnwidth,clip=true]{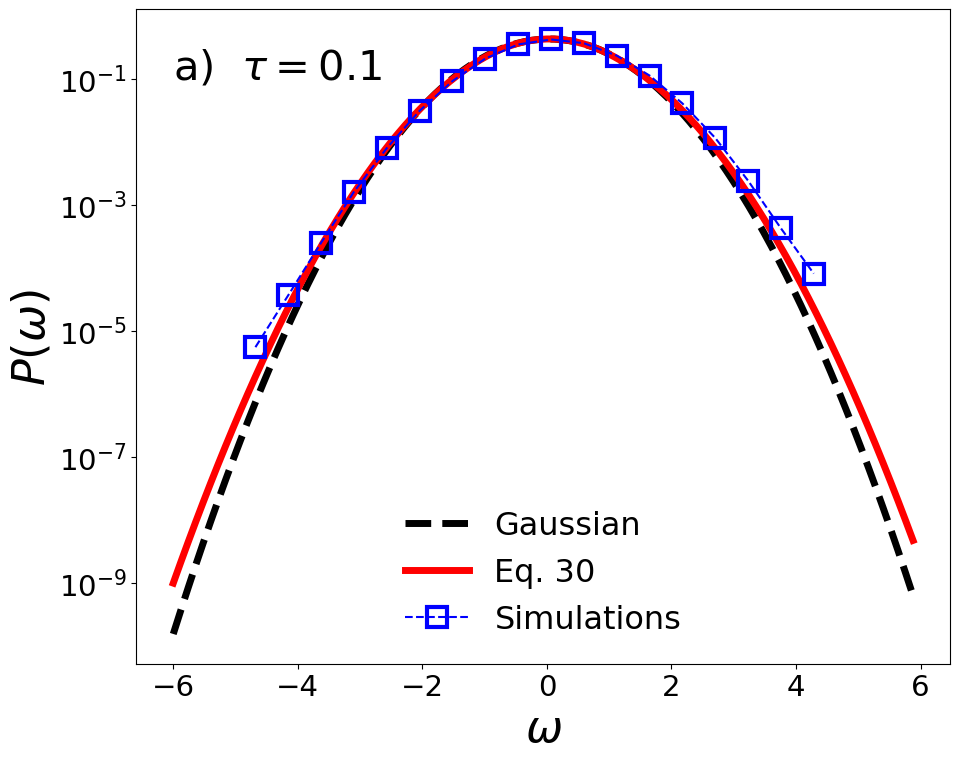}
\includegraphics[width=0.4\columnwidth,clip=true]{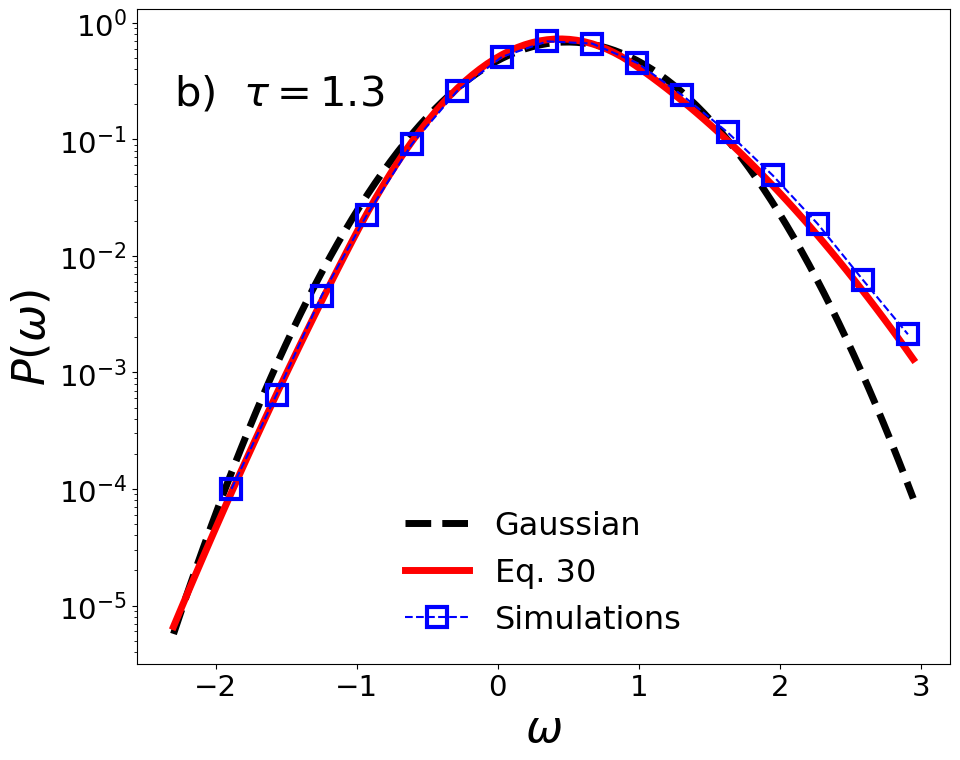}
\caption{Stationary probability distribution computed from Eq. \eqref{eq::Pomega} and from numerical simulations of the model compared with Gaussian distributions having the same mean and variance. Coulomb friction model without memory and $q=1$, $\tau_\mu=0.5$, $\mu=1$, $F_{\text{ext}}=1.3$. For $\tau=0.1$ (panel a), we have a quasi-Gaussian pdf, for $\tau=1.3$ (panel b) deviations from Gaussianity are more evident. 
\label{fig:Pomega}}\end{figure}

For the sake of simplicity, we consider only the Coulomb model without memory, that is Eq.~\eqref{eq::SDEnl_numTest} with $a=0$ and with the force given by Eq.~\eqref{eq::Coulomb}. We analyse the effect of $\tau$ on the diffusion coefficient. As a preliminary analysis, we verified that at small $\tau$, where the force-velocity curve is almost linear, the steady state pdf is quasi-Gaussian, while at large  $\tau$, where the force non-linearity is important, the steady distribution is substantially non-Gaussian (see Fig. \ref{fig:Pomega}).
We also take into account the effect of an external force $F_{\text{ext}}$, which is necessary in order to compare our results with the standard TUR, since its expression requires $F_{\text{ext}} \neq 0$. 
Indeed, contrary to what happens for our bound in Eq. \eqref{eq::nlBound}, where we have simplified the terms $\langle \omega\rangle^2$, in the bound derived from the standard TUR Eq. \eqref{eq::StandardTURbound} the limit $F_{\text{ext}}\to 0$ can be singular depending on the implicit dependence of $\langle \omega\rangle^2$ and $\langle \dot{S}\rangle$ on $F_{\text{ext}}$.

\begin{figure}
\centering
\includegraphics[width=0.4\columnwidth,clip=true]{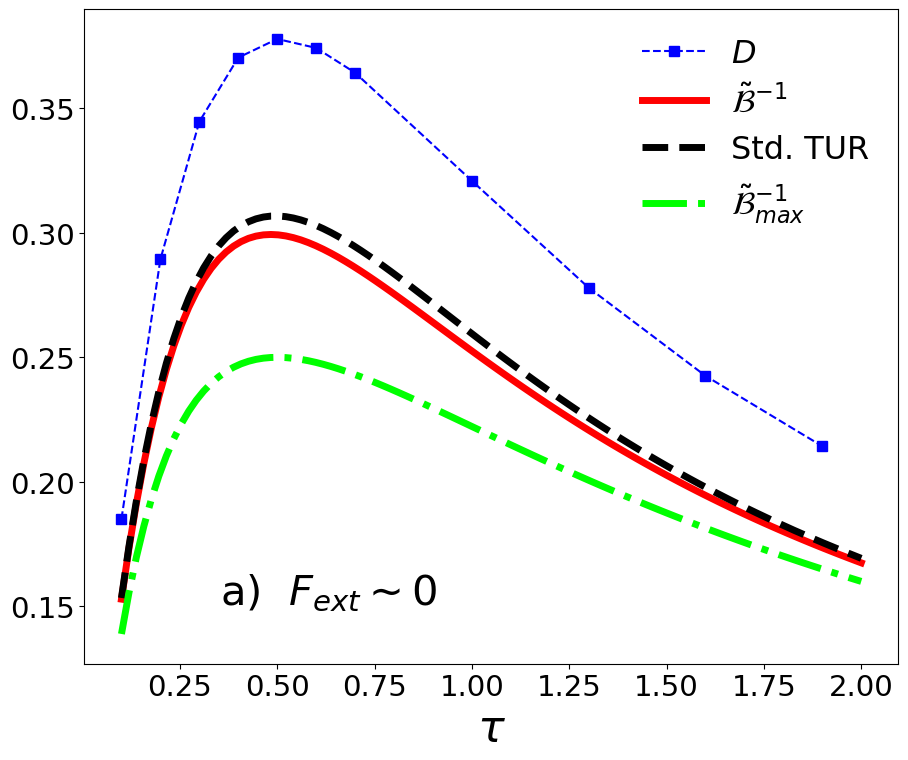}
\includegraphics[width=0.4\columnwidth,clip=true]{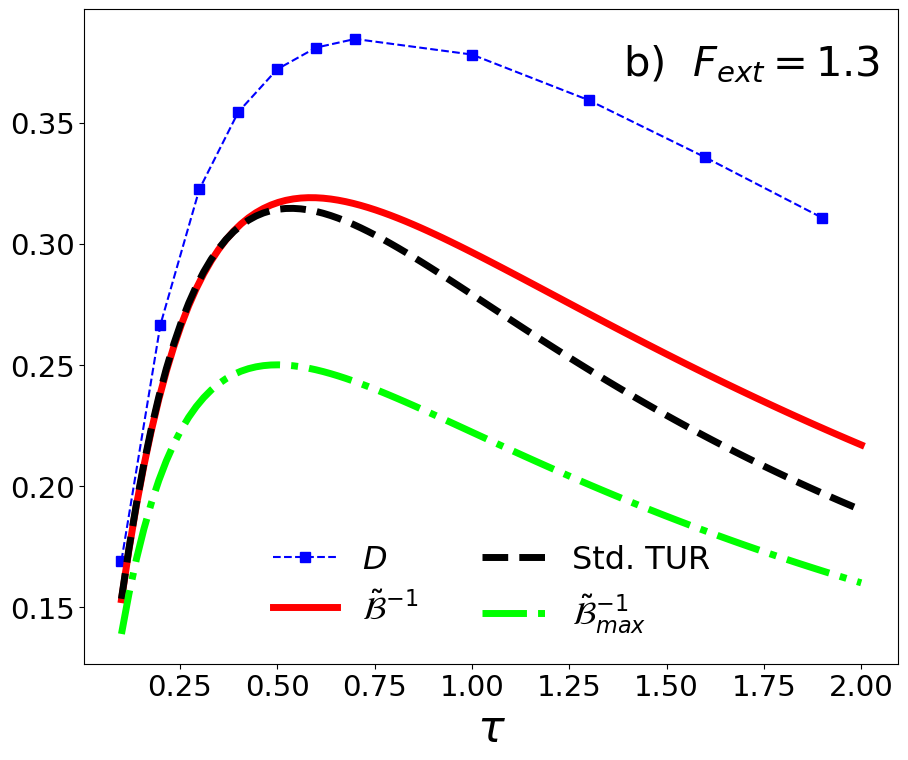}
\caption{Diffusion coefficient and analytical bounds as a function of $\tau$ for the Coulomb friction model without memory and $q=1$, $\tau_\mu=0.5$, $\mu=1$. 
\label{fig:an_bound_tau}}\end{figure}

In Fig.~\ref{fig:an_bound_tau} we show the behavior of the fully analytical bound as a function of $\tau$, for a case with a vanishing external force (panel a) and a non-zero external force (panel b). For this model, the limit $F_{\text{ext} \to 0}$ of $\langle \omega\rangle^2/\langle \dot{S}\rangle$ is regular so we can do the comparison with the bound coming from the standard TUR in panel a. The blue curves correspond to the diffusion coefficient measured from numerical simulations, the black dashed lines to the bound obtained from the standard TUR, the red solid lines correspond to our bound discussed here and the green dash-dotted curves correspond to our bound where $\langle \mathcal{F'}\rangle$ is replaced with $\max(|\mathcal{F}'|)$.
We first note that the non-monotonous behavior in $\tau$ is well captured by all the bounds. Then, we underline that when 
$F_{\text{ext}} \sim 0$ the standard TUR usually surpasses the TUR discussed here. The opposite occurs when an external force is present, particularly the TUR discussed here is largely dominating in the region, at large $\tau$, where non-linear and non-Gaussian effects are important.

\begin{figure}
\centering
\includegraphics[width=0.4\columnwidth,clip=true]{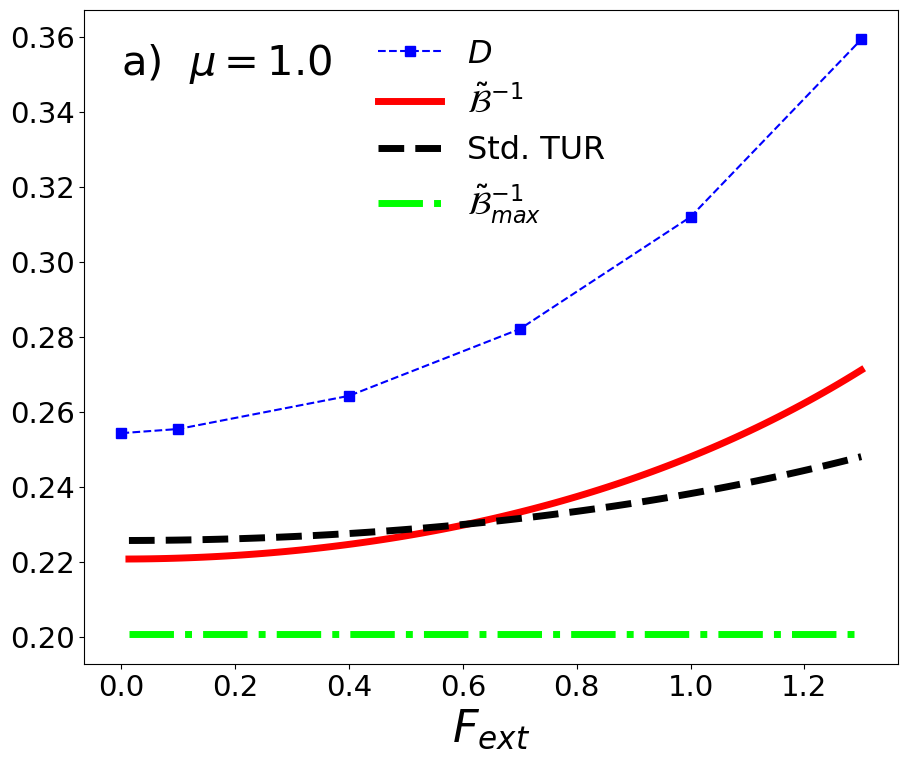}
\includegraphics[width=0.4\columnwidth,clip=true]{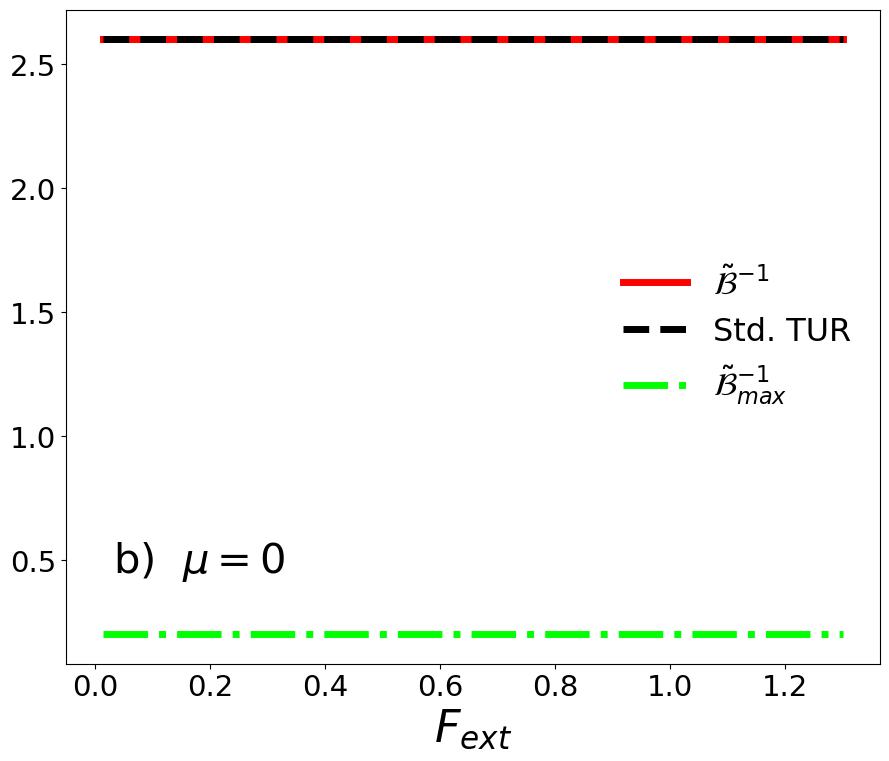}
\caption{Diffusion coefficient and analytical bounds as a function of $F_{\text{ext}}$ for the Coulomb friction model without memory and $\tau=1.3$, $q=1$, $\tau_\mu=0.5$.\label{fig:an_bound_F}}\end{figure}

In Fig.~\ref{fig:an_bound_F} we show the behavior of the fully analytical bound as functions of the external force $F_{\text{ext}}$. In panel a), the effect discussed previously is enhanced: the TUR derived in this paper gives a tighter bound with respect to the one obtained from the standard TUR for $F_{\text{ext}}\gtrsim 0.6$. In panel b, we show the results for regular Brownian diffusion. As we expect, our bound and the one coming from the standard TUR coincide in this case.

\section{Conclusions}

We have reconsidered the information-based derivation of the Thermodynamic Uncertainty Relations, recognising that for a class of Langevin systems with linear coupling and a non-linear friction in the first (velocity) variable, a new thermodynamic uncertainty relation can be obtained. The main application of linearly coupled Langevin equations is for non-Markovian physical models with exponentially decaying memory kernels: each auxiliary variable in the system represents a particular memory time-scale. We remind that it is possible to represent a power-law decaying memory kernel as a sum of a number of exponentially decaying memory kernels (i.e. a number of auxiliary variables), provided that such a number is sufficiently large (typically one for each decade of the power law).

The new uncertainty relation is a bound from below for the mean squared displacement as a function of time. It implies a bound for the diffusion coefficient, a quantity which - in cases with non-linear forces and in particular when more variables are present - is difficult to estimate from the knowledge of the model parameters or from empirical data of other quantities. The analytical bound requires the knowledge of information-theory-related quantities. In order to simplify the expression of the bound, we also show less tight bounds where simpler quantities are required. 

We notice that, in some cases, the standard TUR gives trivial or useless bounds when the external force is absent. 
The new TUR, on the contrary, due to the fact that the entropy production related to the thermal gradients in the system is absent from the denominator, allows one to obtain bounds (not far from the correct value) for the diffusion coefficient also when the external force is absent. When the external force is present we have verified that the TUR derived in this paper can become tighter than the standard one when the system is strongly out of equilibrium.

\section{Acknowledgments}
The authors acknowledge useful discussions with Benjamin Walter. Andrea Puglisi and Alessandro Sarracino also acknowledge the Isaac Newton Institute for Mathematical Sciences for support and hospitality during the programme ‘Mathematics of Movement: an interdisciplinary approach to mutual challenges in animal ecology and cell biology’, when part of the work on this paper was undertaken, supported by the EPSRC Grant Number EP/R014604/1.

\bibliographystyle{abbrv}
\bibliography{biblioTurnew}

\begin{thebibliography}{10}

\bibitem{Barato2015}
A.~C. Barato and U.~Seifert.
\newblock Thermodynamic uncertainty relation for biomolecular processes.
\newblock {\em Phys. Rev. Lett.}, 114:158101, Apr 2015.

\bibitem{baule2010stick}
A.~Baule, H.~Touchette, and E.~Cohen.
\newblock Stick--slip motion of solids with dry friction subject to random
  vibrations and an external field.
\newblock {\em Nonlinearity}, 24(2):351, 2010.

\bibitem{Cantat2013}
I.~Cantat.
\newblock {Liquid meniscus friction on a wet plate: Bubbles, lamellae, and
  foams)}.
\newblock {\em Physics of Fluids}, 25(3):031303, 03 2013.

\bibitem{cerino2015entropy}
L.~Cerino and A.~Puglisi.
\newblock Entropy production for velocity-dependent macroscopic forces: The
  problem of dissipation without fluctuations.
\newblock {\em Europhysics Letters}, 111(4):40012, 2015.

\bibitem{dechant2018multidimensional}
A.~Dechant.
\newblock Multidimensional thermodynamic uncertainty relations.
\newblock {\em Journal of Physics A: Mathematical and Theoretical},
  52(3):035001, 2018.

\bibitem{dieball2023direct}
C.~Dieball and A.~Godec.
\newblock Direct route to thermodynamic uncertainty relations and their
  saturation.
\newblock {\em Physical Review Letters}, 130(8):087101, 2023.

\bibitem{gennes2005brownian}
P.~G.~d. Gennes.
\newblock Brownian motion with dry friction.
\newblock {\em Journal of Statistical Physics}, 119:953--962, 2005.

\bibitem{Gingrich2016}
T.~R. Gingrich, J.~M. Horowitz, N.~Perunov, and J.~L. England.
\newblock Dissipation bounds all steady-state current fluctuations.
\newblock {\em Phys. Rev. Lett.}, 116:120601, Mar 2016.

\bibitem{Hartich2021}
D.~Hartich and A.~c.~v. Godec.
\newblock Thermodynamic uncertainty relation bounds the extent of anomalous
  diffusion.
\newblock {\em Phys. Rev. Lett.}, 127:080601, Aug 2021.

\bibitem{Hasegawa2019II}
Y.~Hasegawa and T.~Van~Vu.
\newblock Uncertainty relations in stochastic processes: An information
  inequality approach.
\newblock {\em Phys. Rev. E}, 99:062126, Jun 2019.

\bibitem{hayakawa2005langevin}
H.~Hayakawa.
\newblock Langevin equation with coulomb friction.
\newblock {\em Physica D: Nonlinear Phenomena}, 205(1-4):48--56, 2005.

\bibitem{horowitz2020thermodynamic}
J.~M. Horowitz and T.~R. Gingrich.
\newblock Thermodynamic uncertainty relations constrain non-equilibrium
  fluctuations.
\newblock {\em Nature Physics}, 16(1):15--20, 2020.

\bibitem{hwang2018energetic}
W.~Hwang and C.~Hyeon.
\newblock Energetic costs, precision, and transport efficiency of molecular
  motors.
\newblock {\em The journal of physical chemistry letters}, 9(3):513--520, 2018.

\bibitem{kim2020learning}
D.-K. Kim, Y.~Bae, S.~Lee, and H.~Jeong.
\newblock Learning entropy production via neural networks.
\newblock {\em Physical Review Letters}, 125(14):140604, 2020.

\bibitem{Lequy2023}
T.~Lequy and A.~M. Menzel.
\newblock Stochastic motion under nonlinear friction representing shear
  thinning.
\newblock {\em Physical Review E}, 108(6), Dec. 2023.

\bibitem{loos2021stochastic}
S.~A. Loos.
\newblock {\em Stochastic systems with time delay: probabilistic and
  thermodynamic descriptions of non-Markovian processes far from equilibrium}.
\newblock Springer Nature, 2021.

\bibitem{manacorda2014coulomb}
A.~Manacorda, A.~Puglisi, and A.~Sarracino.
\newblock Coulomb friction driving brownian motors.
\newblock {\em Communications in Theoretical Physics}, 62(4):505, 2014.

\bibitem{marconi2008fluctuation}
U.~M.~B. Marconi, A.~Puglisi, L.~Rondoni, and A.~Vulpiani.
\newblock Fluctuation--dissipation: response theory in statistical physics.
\newblock {\em Physics reports}, 461(4-6):111--195, 2008.

\bibitem{Muser2011}
M.~H. M\"user.
\newblock Velocity dependence of kinetic friction in the prandtl-tomlinson
  model.
\newblock {\em Phys. Rev. B}, 84:125419, Sep 2011.

\bibitem{plati2023thermodynamic}
A.~Plati, A.~Puglisi, and A.~Sarracino.
\newblock Thermodynamic bounds for diffusion in nonequilibrium systems with
  multiple timescales.
\newblock {\em Physical Review E}, 107(4):044132, 2023.

\bibitem{plyukhin2007nonlinear}
A.~Plyukhin and A.~Froese.
\newblock Nonlinear dissipation effect in brownian relaxation.
\newblock {\em Physical Review E}, 76(3):031121, 2007.

\bibitem{puglisi2009irreversible}
A.~Puglisi and D.~Villamaina.
\newblock Irreversible effects of memory.
\newblock {\em EPL (Europhysics Letters)}, 88(3):30004, 2009.

\bibitem{sarracino2013time}
A.~Sarracino.
\newblock Time asymmetry of the kramers equation with nonlinear friction:
  Fluctuation-dissipation relation and ratchet effect.
\newblock {\em Physical Review E}, 88(5):052124, 2013.

\bibitem{sarracino2013ratchet}
A.~Sarracino, A.~Gnoli, and A.~Puglisi.
\newblock Ratchet effect driven by coulomb friction: The asymmetric rayleigh
  piston.
\newblock {\em Physical Review E}, 87(4):040101, 2013.

\bibitem{seifert2018stochastic}
U.~Seifert.
\newblock Stochastic thermodynamics: From principles to the cost of precision.
\newblock {\em Physica A: Statistical Mechanics and its Applications},
  504:176--191, 2018.

\bibitem{steven1993fundamentals}
M.~K. Steven.
\newblock Fundamentals of statistical signal processing.
\newblock {\em PTR Prentice-Hall, Englewood Cliffs, NJ}, 10(151045):148, 1993.

\bibitem{Hasegawa2019}
T.~Van~Vu and Y.~Hasegawa.
\newblock Uncertainty relations for underdamped langevin dynamics.
\newblock {\em Phys. Rev. E}, 100:032130, Sep 2019.

\bibitem{vanossi2013colloquium}
A.~Vanossi, N.~Manini, M.~Urbakh, S.~Zapperi, and E.~Tosatti.
\newblock Colloquium: Modeling friction: From nanoscale to mesoscale.
\newblock {\em Reviews of Modern Physics}, 85(2):529, 2013.

\end{thebibliography}

\end{document}